\documentclass[graybox]{svmult}

\usepackage{url}
\usepackage{hyperref}
\usepackage{mathptmx}       
\usepackage{helvet}         
\usepackage{courier}        
\usepackage{type1cm}        
\usepackage{wrapfig}                            
\usepackage{csquotes}
\usepackage{mathtools}
\usepackage{makeidx}         
\usepackage{graphicx}        
\usepackage{amsmath}                             
\usepackage{multicol}        
\usepackage[bottom]{footmisc}
\usepackage{slashed}
\usepackage{array}
\usepackage{subcaption}
\usepackage{newtxmath}
\usepackage{float}
\usepackage{graphicx,caption}
\usepackage{geometry}
\makeindex             

\title*{DUNE prospect for leptophobic dark matter}

\author{Sabeeha Naaz, Jyotsna Singh and R. B. Singh}
\institute{Sabeeha Naaz, Dr. Jyotsna Singh and R. B. Singh \at University of Lucknow, Department of Physics,Lucknow-226007,India 
\email{sabeehanaaz0786@gmail.com}
\email{singh.jyotsnalu@gmail.com}
\email{rajendrasinghrb@gmail.com}}
\begin{document}
\maketitle
\vspace{-25mm}
\abstract:{Highly energetic proton/electron beam fixed target experiments extends an opportunity to probe the sub-GeV dark matter and associated 
interactions.
In this work we have explored the sensitivity of DUNE for sub-GeV leptophobic dark matter i.e.  this dark matter barely couples with the leptons. 
Baryon number gauge theory can predicts the existence of a leptophobic cold dark matter particle candidates. In our work, the dark matter candidate 
is considered to be scalar whose mass is defined by the symmetry breaking of new baryonic gauge group $U(1)_{B}$. In this scenario a
light scalar dark matter couples with the standard model candidates via vector boson mediator $V_{B}$ which belongs to the baryonic gauge group 
$U(1)_{B}$.
This leptophobic dark matter dominantly couples to the quarks. Under this scenario new parameter space for $\alpha_{B}$ is explored by DUNE
for leptophobic dark matter candidates.
This new parameter space allowed $\alpha_{B}$ to get lower value than the present exiting constraint value of $\alpha_{B}$ i.e. $10^{-6}$.

\section{Introduction:}

Existence of dark matter (DM) particle is supported by many gravitational phenomena in astrophysics and cosmology but the nature of the dark matter
is still to be discovered. These gravitational evidences motivate the physicist to know about the dark matter (DM) particle candidates and their 
interactions with ordinary matter if any. None of the running experiments are able to probe the DM candidates hence different theoretical assumptions or
models are floating in scientific community to describe their nature. DM candidate are assume to be a non-baryonic, weakly interacting and stable.
Further these DM candidates
are assumed to be weakly interacting and stable. 

On the basis of DM mass and their velocities, DM candidates can be broadly classified into two categories; hot and cold dark matter.
All relativistic and super-relativistic DM particles are called hot DM whereas all non-relativistic DM particles are known as cold DM. Most part of the
DM is cold because they move with non-relativistic speed and they are capable of forming small structure galaxies rather than massive galaxies.
While hot DM can not form small structure of galaxies by the free streaming process. As we observe
small structure galaxies hence it is assumed that DM is a mixture of small amount of hot DM with predominant cold DM.
Different DM physics models needs different experimental environment for the detection of DM. 
The experimental techniques proposed for dark matter detection can be broadly classified into three categories.
\begin{itemize}
 \item \textbf{Direct detection:} In the direct detection technique we suggest that the DM candidates can be detected by the measurement of recoil 
 energy of nucleons but this detection technique becomes less sensitive for the detection of sub-GeV dark matter because in this mass range of DM
 the recoil energy of nucleons
 decreases below the detector threshold value
 of the detector.
 \item \textbf{Indirect detection:} In indirect detection, the standard model (SM) particles produced via annihilation of DM particles are studied.
 \item \textbf{Collider technique:} Beam collider technique looks for the missing transverse energy in a event, for the DM detection.
\end{itemize}
 
 In this work, we have focused on the MeV to GeV mass range of DM because a lot of work have been done for mass range 1 GeV to 10 TeV [WIMP
 (weakly interactive massive particles) as a DM candidate] at direct detection experiments but these experiments have estimated null results for WIMP
 \cite{79, 80, 81, 82, 83}.  
 Hence sub-GeV mass DM particles can be a potential candidates of DM family and
 the sub-GeV DM candidates can be explored by direct detection, in neutrino laboratories. 
 To improve the sensitivity of DM candidates fixed target approach is considered in direct detection experiment but
 one major drawback of  this approach is the presence of huge neutrino
 background. These particles (neutrino) can mimic the DM signatures hence for the detection of DM neutrino background reduction becomes necessary.
 To mitigate the neutrino background in our DM detection work, 
 we have considered DUNE experiment running in beam dump mode.
 Fir the first time, neutrino experiment in the beam dump mode was carried by MiniBooNE in 2014 \cite{69,8}.
 The fixed target experiment in beam dump mode opens the door for the study of DM and hidden sector physics via different DM portals.
 
 A large number of DM studies \cite{69, 66, 67, 68, 70, 71, 72, e} have focused on the kinetic mixing scenario, in which DM of hidden sector 
 couples with the SM particles via kinetic mixing parameter $(\epsilon)$; which kinetically mixes the dark photon $(\gamma_{D})$ mediator of hidden sector
 and ordinary photon. This is one of the
 possible way of interaction of DM with SM particles but there are other large number of scenarios of DM interaction with SM particles.
 These scenarios needs to be probed for imposing better constraints on DM parameter space.
 In one of this scenario, DM candidates of hidden sector interacts with the SM particles via a new massive vector gauge boson mediator
 $(V_{B})$
 of new baryonic gauge group $U(1)_{B}$. This vector gauge boson $V_{B}$
 dominantly couples with the quarks and is called \enquote{leptophobic dark matter}.
 
 In our work, using fixed target experiment in beam dump mode we have estimated the sensitivity of DUNE (Deep underground neutrino) experiment \cite{1}
 for capturing the signatures of leptophobic dark matter. A highly energetic proton beam of energy 120 GeV is used to produce a
 boosted dark matter beam through vector boson mediator $V_{B}$. Here three modes of DM production are taken into consideration in our analysis. 
 These modes are pseudosmesons decay, 
 bremsstrahlung and parton-level production channels. The DM signatures are captured in
 near DUNE detector by looking at DM-nucleon elastic scattering.
                
 \vspace{-3mm}
\section{Leptophobic Dark Matter model:}
The accurate measurements of CMB (Cosmic Microwave Background) provides an important information regarding the DM beyond its gravitational interactions.
In large class of models where DM is considered to be thermal relic, its cosmological abundance is calculated by correct thermal relic density of DM
($\sim$ 22\% of the energy density of the universe) which can be obtained by the measurement of s-wave annihilation cross-section 
$\langle\sigma_{s}\rangle \sim 3\times10^{-26} cm^{3}/s$ during freeze-out. In our studies we are interested in DM candidates whose masses are below 
a few GeV, hence bounds
imposed by the direct detection will be very mild. The most stringent constraints on these DM candidates can be imposed by the bounds on energy injection 
around redshift
$z \sim 100 - 1000$, coming from the CMB.
The s-wave annihilation of DM into charged SM particles, in particular rules out DM masses below 10 GeV but by adding p-wave process into 
annihilation cross-section of DM, the DM of masses less than 10 GeV can be produced.
For the correct reproduction of DM annihilation cross-section with only s-wave process, the most viable annihilation mode of DM is, its annihilation 
into neutrino like states. 
In leptophobic DM model, the DM can be a viable thermal relic DM candidate only if the DM annihilation occurs into the light baryonic neutrinos via 
s-wave process.
In this annihilation process because of weak interaction of baryonic neutrinos with matter they are unable to ionize the hydrogen and helium gases
hence the problem of energy injection around redshift $z \sim 100 - 1000$ are
completely ignore, during and after the recombination. 
In this scenario the required annihilation cross-section of thermal relic DM (of $\sim pb$ order) can be achieved.
The required value of annihilation cross-section imposes constraints on the on-shell production mode and
off-shell production mode of DM candidates. In off-shell production mode ($m_{V_{B}} < m_{\chi}$; $m_{V_{B}}$ is the mass of vector boson 
mediator $V_{B}$ and
$m_{\chi}$ is the DM mass), $\chi\chi^{\dagger} \to V_{B}^{*} \to \nu_{b}\bar{\nu_{b}}$
annihilation process can achieve thermal relic DM annihilation cross-section $\langle\sigma v\rangle \sim 1 pb$ by imposing
$\alpha_{B}^{2} \sim 10^{-11}(m_{\chi}/100 MeV)^{2}$ bound. Whereas for the on-shell production mode ($m_{V_{B}} > m_{\chi}$), baryonic
fine structure constant $\alpha_{B}$ would require slightly larger value.
In our work we have considered, on-shell and off-shell production modes for the production of DM. 
In both
production modes baryonic constant $G_{B} = 4\pi\alpha_{B}/m_{V_{B}}^{2}$ would necessarily be greater than the weak Fermi constant $G_{F}$,
$G_{B} \sim (10^{2} - 10^{3}) \times G_{F}$ \cite{A}.

A benchmark model for sub-GeV leptophobic scalar DM candidates which are charged under a new baryonic gauge group $U(1)_{B}$,
couples with the SM particles through a new vector boson mediator $V_{B}$ which dominantly interacts with quarks. 
We have used generation independent coupling of $V_{B}$ with quarks to make model simple as it avoids the tree level flavor 
changing neutral current interactions. In this consideration, to allow the renormalizable Yukawa couplings of quarks to the Higgs boson of SM,
coupling of $V_{B}$ with right and left handed quarks should be the same. 

Our considered model, due to the addition of new gauge group $U(1)_{B}$ to the standard model gauge group suffers from gauge anomalies and can be
considered as
non-renormalizable effective field theory with a cutoff $\Lambda_{UV}$ \cite{q}. New states, either at or below this cutoff must be introduced
for the theory to remain consistent. The simplest choice for addition of new states are new chiral fermion which help in the cancellation
of the anomaly.
Here in this work
our focus is on GeV scale phenomenology hence the exact details of the UV completion can be ignored and we can focus our attention on low energy effective
field theory of a local $U(1)_{B}$ symmetry under which the DM is charged.
The Lagrangian of leptophobic scalar DM for low energy effective field theory can be expressed as,
\begin{equation}
 \mathcal{L}_{DM} = |D_{\mu}\chi|^{2} - m_{\chi}^{2}|\chi|^{2} - \frac{1}{4}(F_{B}^{\mu\nu})^{2} + \frac{1}{2}m_{V_{B}}^{2}(F_{B}^{\mu})^{2} - \frac{\epsilon}{2}F_{B}^{\mu\nu}
 F_{\mu\nu} + g_{B}F_{B}^{\mu}J_{B\mu} + ......
\end{equation}


where $F_{\mu\nu} = [\partial_{\mu}F_{\nu} - \partial_{\nu}F_{\mu}]$ is the field of SM,
$F_{B}^{\mu\nu} = [\partial^{\mu}V_{B}^{\nu} - \partial^{\nu}V_{B}^{\mu}]$ is field strength of $U(1)_{B}$,
$D = \partial - ig_{B}q_{B}V_{B}$, $g_{B}$ is baryonic gauge coupling and $q_{B}$ is baryonic charge of $U(1)_{B}$. The $J_{B}^{\mu}$
represents sum of baryonic current over all quarks species i.e. $J_{B}^{\mu} = \frac{1}{3}\sum_{i}\bar{q_{i}}\gamma^{\mu}q_{i}$. Above Lagrangian includes
baryonic coupling ($g_{B}$) scenario as well as kinetic mixing ($\epsilon$) interaction scenario.
Two different scenarios at a time are possible and may give some interesting results,
but for most cases either the baryonic coupling scenario or the kinetic mixing scenario will dominate.
Therefore for leptophobic DM scenario we have set the value of kinetic mixing parameter $\epsilon \to 0$ as we want to check constraint on DM
parameter space in the presence of baryonic coupling.

\section{Production of Leptophobic Dark Matter:}
We have focused on three production channels of leptophobic dark matter for DUNE experiment. The considered production modes are as follows,
\begin{itemize}
 \item Pseudoscalar mesons decay
 \item Bremsstrahlung process
 \item Parton-level production
\end{itemize}
\subsection{Pseudoscalar Mesons Decay}
This production channel dominates over all production modes of dark matter at lower masses of vector boson mediator $V_{B}$. In this mode
vector boson mediator $V_{B}$ are produced via radiative decay of pseudoscalar mesons $\varrho = \pi^{0}, \eta$ \cite{66}\cite{67}.
These secondary mesons are produced from the primary interaction of $p(p)$ and $p(n)$.

\begin{equation}
 p + p(n) \to X + \pi^{0}/\eta \to X + \gamma + V_{B} \to X + \gamma + \chi + \chi^{\dagger}
 \label{1}
\end{equation}

The pseudoscalar mesons coupling with vector boson $V_{B}$ takes place under the gauged Wess-Zumino-Witten (WZW) lagrangian \cite{b,c,d}.
If the mass of secondary meson is greater than the mass of the vector boson mediator $V_{B}$ i.e. $m_{V_{B}} < m_{\varrho}$ 
or if the mass of
DM particle produced is such that $2m_{\chi} < m_{V_{B}} < m_{\varrho}$ then $V_{B}$ will be produced on-shell and further it will 
decay into a pair of DM candidates. 
Branching ratio of pseudoscalar mesons decay to dark matter particles is calculated using narrow width approximation \cite{67} which is equal 
to the product of 
mesons decay to $V_{B}$ and decay of $V_{B}$ to dark matter. 

\begin{equation}
Br(\varrho \to \gamma\chi\chi^{\dagger}) = Br(\varrho  \to \gamma V_{B}) Br(V_{B} \to \chi\chi^{\dagger}) \\
\label{2}
\end{equation}
\begin{equation}
Br(\varrho \to \gamma V_{B}) = 2\left(c_{\varrho}\frac{g_{B}}{g} - \epsilon\right)^{2} \left(1 - \frac{m_{V_{B}}^{2}}{m_{\varrho}^{2}}\right)^{3} 
 Br(\varrho \to \gamma \gamma)
 \label{3}
\end{equation}

Where $g_{B}$ is baryonic coupling constant, $g$ is electromagnetic coupling constant, $m_{\varrho}$ is mass of the pseudoscalar mesons and 
the value of $c_{\varrho}$ is different for different mesons i.e. $c_{\pi^{0}} = 1$ and $c_{\eta} \approx 0.11$ \cite{B}.
We have used the BMPT (Beryllium Material Proton Target) distribution fits \cite{54} in rejection sampling
to simulate the momentum and angular distribution of mesons.
For the detailed calculation check the reference \cite{a}.

\subsection{Bremsstrahlung Process}
For intermediate masses of vector boson mediator $V_{B}$, resonant vector mesons $R = \omega$ decay by the bremsstrahlung process 
dominates over other decay processes.
Here the mass of $V_{B}$ is close to the mass of resonant vector mesons. 

\begin{equation}
p + p(n) \to p + p(n) + V_{B} \to .... + \chi + \chi^{\dagger}
 \label{4}
\end{equation}

In this process vector mesons mixes with vector boson mediator $V_{B}$. 
This process generates a nearly collimated beam of vector boson $V_{B}$. The four momentum of incident proton of mass $m_{p}$ is 
$q = (E_{p}, 0, 0, Q)$ where $E_{p} = Q + \frac{m_{p}^{2}}{2Q}$. The four momentum of outgoing vector boson mediator of mass $m_{V_{B}}$ is 
$q_{V_{B}} = (E_{V_{B}}, q_{\perp}\cos(\phi),
q_{\perp}\sin(\phi), Q.z)$ 
where $E_{V_{B}} = Q.z + \frac{q_{\perp}^{2} + m_{V_{D}}^{2}}{2Q.z}$, $Q.z = q_{\parallel}$ and z is a fraction of proton beam momentum 
carried away by outgoing vector boson $V_{B}$ in the direction of proton beam. Here $q_{\perp}$ and $q_{\parallel}$ are transverse and longitudinal 
components 
of the momenta of $V_{B}$ .\\
By Weizs$\ddot{a}$cker-Williams approximation the rate of production of vector boson $V_{B}$ production per proton is as follows \cite{56,57,e},\\
\begin{equation}
 \frac{d^{2}N_{V_{B}}}{dz dq_{\perp}^{2}} = 
 \frac{\sigma_{pA}[2m_{p}(E_{p} - E_{V_{B}})]}{\sigma_{pA}(2m_{p}E_{p})}F^{2}_{1, N}(q^{2})f_{yx}(z, q_{\perp}^{2})
 \label{14}
\end{equation}
\\
here $\sigma_{pA} = f(A)\sigma_{pp}$, $f(A)$ is a function of atomic number $A$ and $f_{yx}(z, q_{\perp}^{2})$
is a splitting weight-function of photon which relates before and after differential scattering cross section \cite{56},\\

$f_{yx}(z, q_{\perp}^{2}) = \frac{\alpha_{B}}{2\pi H}\left[\frac{1 + (1 - z)^{2}}{z} - 2z(1-z)\left
(\frac{2m_{p}^{2} + m_{V_{B}}^{2}}{H} - z^{2}\frac{2m_{p}^{4}}{H^{2}}\right)
+ 2z(1 - z)(z + (1 - z)^{2})
\frac{m_{p}^{2}m_{V_{B}}^{2}}{H^{2}} + 2z(1 - z)^{2}\frac{m_{V_{B}}^{4}}{H^{2}} \right]$ 
\\
here $H = q_{\perp}^{2} + (1 - z)m_{V_{B}}^{2} + z^{2}m_{p}^{2}$.

Since radiative $V_{B}$ has time-like momentum and
time-like form factor $F_{1,N}(q^{2})$ expresses off-shell mixing of vector bosons with vector mesons in appropriate kinematic region.
Baryonic vector portal considers both; proton form factor $F_{1,p}(q^{2})$ and neutron form factor $F_{1,n}(q^{2})$ \cite{58}.
These incorporates only isoscalar Breit-Wigner components \cite{58} $\omega$-like in the spacelike regime
and this form factor is not completely resolved for $\omega$. 
Above 1 GeV, the form factors suppresses the rate of production of virtual bosonic mediator hence above this energy direct parton level
production dominates over other channels.

To calculate the dark photon production rate, equation [\ref{14}] must be integrated over $p_{\perp}$ and $z$ in a range that satisfies 
some kinematic conditions \cite{56} expressed as,
\begin{equation}
 E_{p}, E_{V_{B}}, E_{P} - E_{V_{B}} \gg m_{p}, m_{V_{B}}, \left|{q_{\perp}}\right|
\end{equation}
A range $z \in [0.2,0.8]$ and $\left|{p_{\perp}}\right| = 0.4$ for DUNE is selected as is satisfies the above kinematic conditions.

\subsection{Parton-Level Production of Dark Matter}

Above 1 GeV of vector boson mediator $V_{B}$ mass, this channel becomes significant. This process works under the narrow width approximation via
$q\bar{q} \to V_{B}$ and can be written as,

\begin{equation}
 p + p(n) \to X + V_{B} \to X + \chi\chi^{\dagger}
 \label{5}
\end{equation}

Dark matter pair production cross-section at parton-level can be expressed as,

\begin{equation}
 \sigma(pp(n) \to X + V_{B} \to X + \chi\chi^{\dagger}) = \sigma(pp(n) \to V_{B}) Br(V_{B} \to \chi\chi^{\dagger})
 \label{6}
\end{equation}
 where $\sigma(pp(n) \to V_{B})$ cross-section for the production of vector boson $V_{B}$ and can be written as,
\begin{equation}
 \sigma(pp(n) \to V_{B}) = 
 \frac{\pi}{3m_{V_{B}}^{2}}
 \sum_{q}\left(\frac{g_{B}}{3} - \epsilon gQ_{q}\right)^{2} \int_{\zeta}^{1} \frac{dx}{x} \tau \left[f_{q/p}(x)f_{\overline{q}/p(n)}(\frac{\zeta}{x}) 
 + f_{\overline{q}/p}(x)f_{q/p(n)}(\frac{\zeta}{x})\right]
 \label{7}
\end{equation}
where $\zeta = m_{V_{B}}^{2}/s$, $\sqrt{s}$ is the hadron-level
center of mass energy and $Q_{q}$ is quark charge in the unit of positron electric charge. 
To calculate the cross-section of the DM production, we have used CTEQ6.6 PDFs \cite{60} and have set $Q = m_{V_{B}}$ which is allowed to vary from
$m_{V_{B}}/2$ to $2m_{V_{B}}$.
In above equation $f_{q/p(n)}(x)$ is the parton distribution function (PDF) which gives the probability of extraction of
quarks and gluons with longitudinal momentum fraction $x$ from a proton (neutron).
Details of the cross-section calculation are discussed in the references \cite{68, e, a}.

\begin{figure}[H]
\centering
\includegraphics[width=0.8\linewidth]{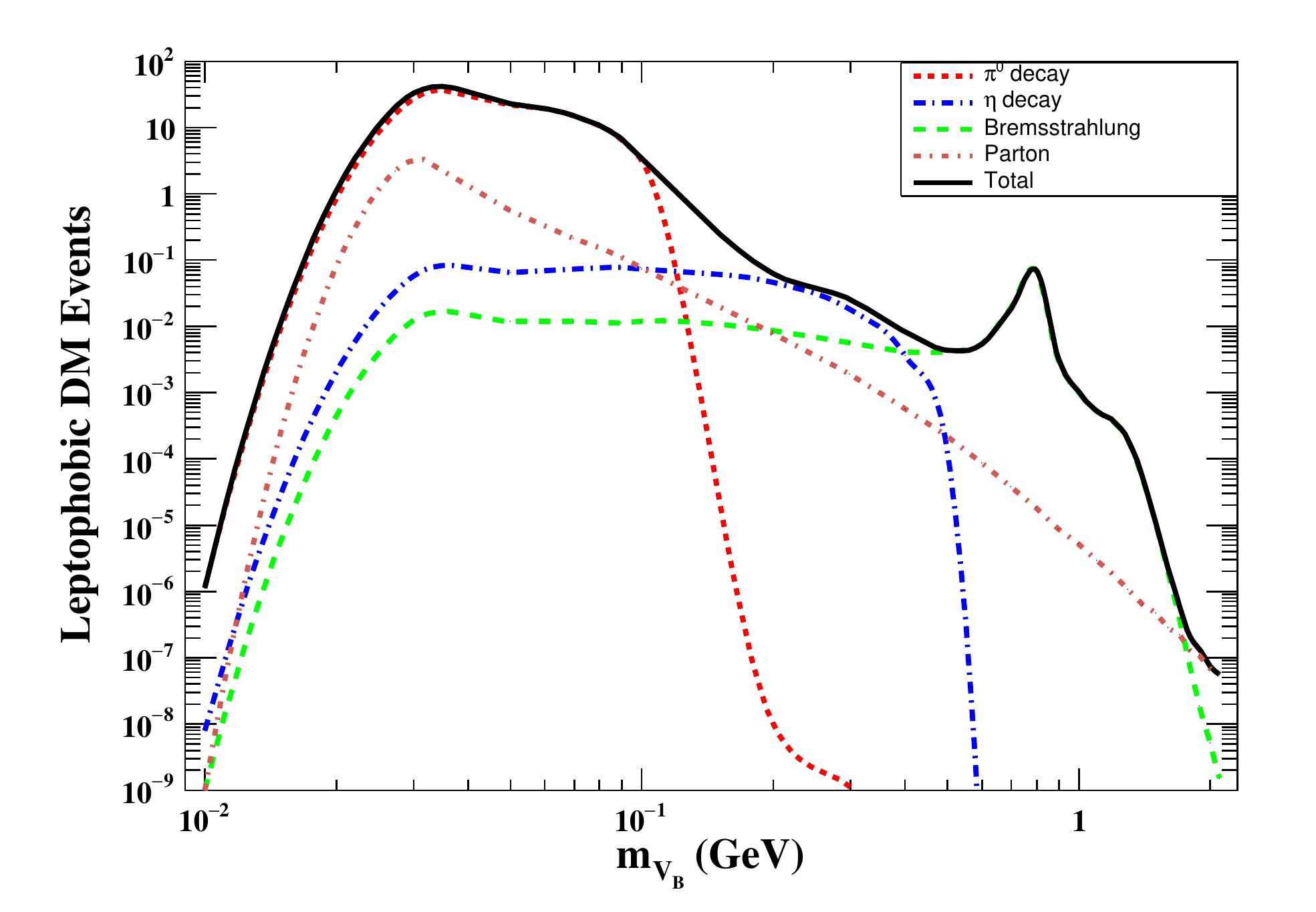}
 \caption{Dark matter-nucleon scattering event plot with the variation of mediator vector boson mass for all distinct channels. 
 Here $m_{\chi} = 0.01$ GeV, $\epsilon$ = $0$, $\alpha_{B} = 10^{-6}$ and POT = $1.1 \times 10^{21}$.}
\label{fig2}
\end{figure}

\section{Scattering Cross-Section of Leptophobic Dark Matter:}

In the considered leptophobic dark matter model, dark matter dominantly couples with the quarks whereas it do not couple with the leptons.
In our work we have focused on the
neutral current elastic scattering of dark matter with nucleons present in the DUNE near detector. The differential cross-section of 
neutral current DM-nucleon 
elastic scattering which is similar to the neutrino-nucleon neutral current scattering \cite{f, g} can be expressed as,

\begin{equation}
 \frac{d\sigma_{\chi N\to \chi N}}{dE_{\chi}} = \alpha_{B}q_{B}^{2} \times 
\frac{\tilde{F}_{1,N}^{2}(Q^{2})A(E,E_{\chi}) + \tilde{F}_{2,N}^{2}(Q^{2})B(E,E_{\chi}) + \tilde{F}_{1,N}(Q^{2})\tilde{F}_{2,N}(Q^{2})C(E,E_{\chi})}
{(m_{\gamma_{D}}^{2} + 2m_{N}(E - E_{\chi}))^{2}(E^{2} - m_{\chi}^{2})}
\label{8}
\end{equation}

where $E$ and $E_{\chi}$ represents energy of the incoming and outgoing DM, $m_{N}$ is the mass of nucleons $(N = p, n)$ 
and $Q^{2} = 2m_{N}(E - E_{\chi})$
is the momentum transfer. The $\tilde{F}_{1,N}^{2}(Q^{2})$ and $\tilde{F}_{2,N}^{2}(Q^{2})$ are monopole and dipole form factors \cite{a} and the 
value of kinematic functions for complex scalar DM are listed as,

$ A = 2m_{N}EE_{\chi} - m_{\chi}^{2}(E - E_{\chi});$ \\

$ B = \frac{1}{4}(E - E_{\chi})[(E + E_{\chi})^{2} - 2m_{N}(E - E_{\chi}) -4m_{\chi}^{2}];$ \\

$ C = -(E - E_{\chi})(m_{N}(E - E_{\chi}) + 2m_{\chi}^{2})$.

\section{Signal Rates:}

To simulate the DM event rates its essential to incorporate the relevant cuts of detector geometry and energy resolution
in the simulation tool. In this work the detector limitations are taken from the reference \cite{e}. 
The expression for the DM events produced by the pseudoscalar mesons and vector mesons can be stated as,

\begin{equation}
 N_{\chi N \to \chi N} = n_{A}\epsilon_{eff} \sum_{M=\pi^{0},\eta,\omega}\left[N_{M}Br(M \to V_{B} + ...)Br(V_{B} 
 \to \chi\chi^{\dagger}) \times \left(\frac{1}{N_{\chi M}} \sum_{i} L_{i} \sigma_{\chi N, i}\right)\right]
 \label{10}
\end{equation}

where $n_{A}$ is atomic number density of detector material, $\epsilon_{eff}$ is detector efficiency, $N_{M}$ is total number of mesons produced in the 
target, $N_{\chi M}$ is total number of DM trajectories produced by relevant production channels, $L_{i}$ is the length of DM trajectory in the detector.
The DM-nucleon elastic scattering cross-section $\sigma_{\chi N}$ is defined as,

\begin{equation}
 \sigma_{\chi N}(E) = \int_{E_{\chi}^{min}}^{E_{\chi}^{max}} dE_{\chi} \sum_{N=p,n} f_{N}
 \frac{d\sigma_{\chi,N}}{dE_{\chi}}
 \label{11}
\end{equation}
where $E$ represents the energy of incoming dark matter, $E_{\chi}$ represents the energy of outgoing dark matter and $E_{\chi}^{min/max}$ is minimum 
and maximum energy of outgoing DM which is calculated by the relevant
experimental cuts that are derived by the experimental data of nucleon recoil momentum $q$ ($q$ = $\sqrt{2m_{N}(E - E_{\chi})}$). 
In baryonic vector portal we take $f_{p,n}$ = A (A is atomic number) for the elastic or quasi-elastic scattering of DM. For parton-level production
channel we have substituted $N_{V_{B}}$ (total number of produced vector bosons) in place of $N_{M}Br(M \to V_{B} + ...)$ in above equation [\ref{10}].
Total DM events are evaluated by adding the DM events produced via three different channels considered in this work.

\section{Constraints on the Leptophobic Dark Matter:} 

In this paper we are checking the sensitivity of DUNE fixed target experiment in beam dump mode for leptophobic DM. 
Several essential constraints that needs to be considered for checking the DUNE sensitivity for leptophobic DM are listed below in brief.

\begin{itemize}
 \item \textbf{Direct Detection:} Direct detection experiments probe the cross-section of DM-nucleon elastic scattering.
 The CRESST-II \cite{80}
 experiment provides best limit on the recoil energy of nucleons. The CRESST-II experiment can explore the sensitivity of DM masses
 below 0.5 GeV with detection threshold of nuclear recoil 307 eV. The DM-nucleon scattering cross-section for baryonic current can be expressed as,
 
 \begin{equation}
  \sigma_{\chi N} \sim \frac{16\pi\alpha_{B}^{2}\mu_{\chi,N}^{2}}{m_{V_{B}}^{4}}
 \end{equation}

 where, $\mu_{\chi,N}$ is the reduced mass of DM and nucleons. 
 
 \item \textbf{Constraints on masses of the vector boson mediator $V_{B}$:} 
 The CDF (Collider Detector at Fermilab) collaboration imposes a rigorous constraints on monojet, 
 $pp \to $jet + missing energy. The limit imposed on the quarks coupling $g_{u} < 0.026$ and $g_{d} < 0.04$ are largely independent 
 of vector boson mass $m_{V_{B}} \textless 10 GeV$ \cite{h, i}.
 \item \textbf{$\pi^{0} \to \gamma + invisible$:} The Brookhaven alternating gradient synchrotron \cite{j} imposes limit on the branching ratio 
 $Br(\pi^{0} \to \gamma V_{B}) < 5 \times 10^{-4}$.
 \item \textbf{$K^{+} \to \pi^{+}\nu\bar{\nu}$:} The $K$-decay imposes a limits on the branching ratio of $Br(K^{+} \to \pi^{+} V_{B}) < 10^{-6}$ for 
 $m_{V_{B}}$ = 1.8 MeV and $Br(K^{+} \to \pi^{+} V_{B}) < 7 \times 10^{-7}$ for $m_{V_{B}}$ = 100 MeV \cite{k, l}. 
 \item \textbf{$J/\psi \to invisible$:}
 BES (Beijing Spectrometer) collaboration imposes a constraint on the branching ratio
 $Br(J/\psi \to invisible) < 7\times10^{-4}$ for larger value of $m_{V_{B}}$ \cite{o}. 
 \item \textbf{Angular Dependence in Neutron Scattering:} 
 The constraints imposed by neutron scattering on the baryonic fine structure constant $\alpha_{B}$
 ($\alpha_{B} = \frac{g_{B}^{2}}{4\pi}$)
 which couples the DM candidates and SM particles for mediator vector boson mass $m_{V_{B}} > 1$ MeV is expressed below \cite{m, n},
 \begin{equation}
  \alpha_{B} < 3.4 \times 10^{-11} \left(\frac{m_{V_{B}}}{MeV}\right)^{4}
 \end{equation}

\end{itemize}
Existing constraints on the leptophobic DM model are shown in figure [\ref{fig1}] and [\ref{fig3}]. The plot [\ref{fig1}] shows the constraints on 
the $U(1)_{B}$ model in the $\alpha_{B} - m_{V_{B}}$ parameter space while the plot [\ref{fig3}] shows the constraints on the $U(1)_{B}$ model in 
the $\alpha_{B} - m_{\chi}$ parameter space for 10 MeV DM mass and $\epsilon = 0$. The other constraints plotted in figure [\ref{fig1}] and [\ref{fig3}]
are taken from
$K^{+} \to \pi^{+}\nu\bar{\nu}$,
$\pi^{0} \to \gamma + invisible$, Monojet(CDF), Neutron Scattering and $J/\psi \to invisible$ experiments.

\section{Simulation and Results:}

We have used BdNMC (Beam dump Neutrino Monte Carlo) \cite{61} simulation tool to compute the sensitivity of DUNE detector for leptophobic DM. This
simulation tool works on the direct detection technique to probe the sensitivity of the considered experiment running in beam dump mode.

The $120 GeV$ proton beam of DUNE experiment produces charged and neutral mesons after hitting the beryllium target. 
These mesons are allowed to propagate through the decay pipe which is made up of steel. The neutral mesons quickly decays in the decay pipe while charged
mesons propagate further and get absorbed in the decay pipe before they decay. 
The neutral mesons decay into vector mediator $V_{B}$ which further decays into a pair of DM candidates in the center of mass frame of $V_{B}$.
The total DM particles produced via distinct production channels 
(visit section[3]) reaches the DUNE near detector which is a cylinder of length $5$ m, diameter $5$ m and $1$ ton ($Ar:CH_{4}::90:10$) fiducial mass 
\cite{63}.
Using BdNMC simulation tool the DM event rate are estimated by imposing
production distribution function $f(p, \theta)$, specific geometric cuts \cite{63} and recoil energy cuts on nucleons 
$E_{R} \in [0.1,2]$ in equation \ref{10}.
The trajectory of each DM particle which intersects the fiducial mass
of near detector are recorded in the form of the energy of DM candidates.  

In our work, the sensitivity of fixed target DUNE experiment in beam dump mode is explored for leptophobic DM at DUNE near detector.
In the considered DM model,
DM candidates couples
with the SM particles via vector boson mediator $V_{B}$ of a new baryonic gauge group $U(1)_{B}$ with coupling strength $\alpha_{B}$. 

\begin{figure}[H]
\centering
\includegraphics[width=1.0\linewidth]{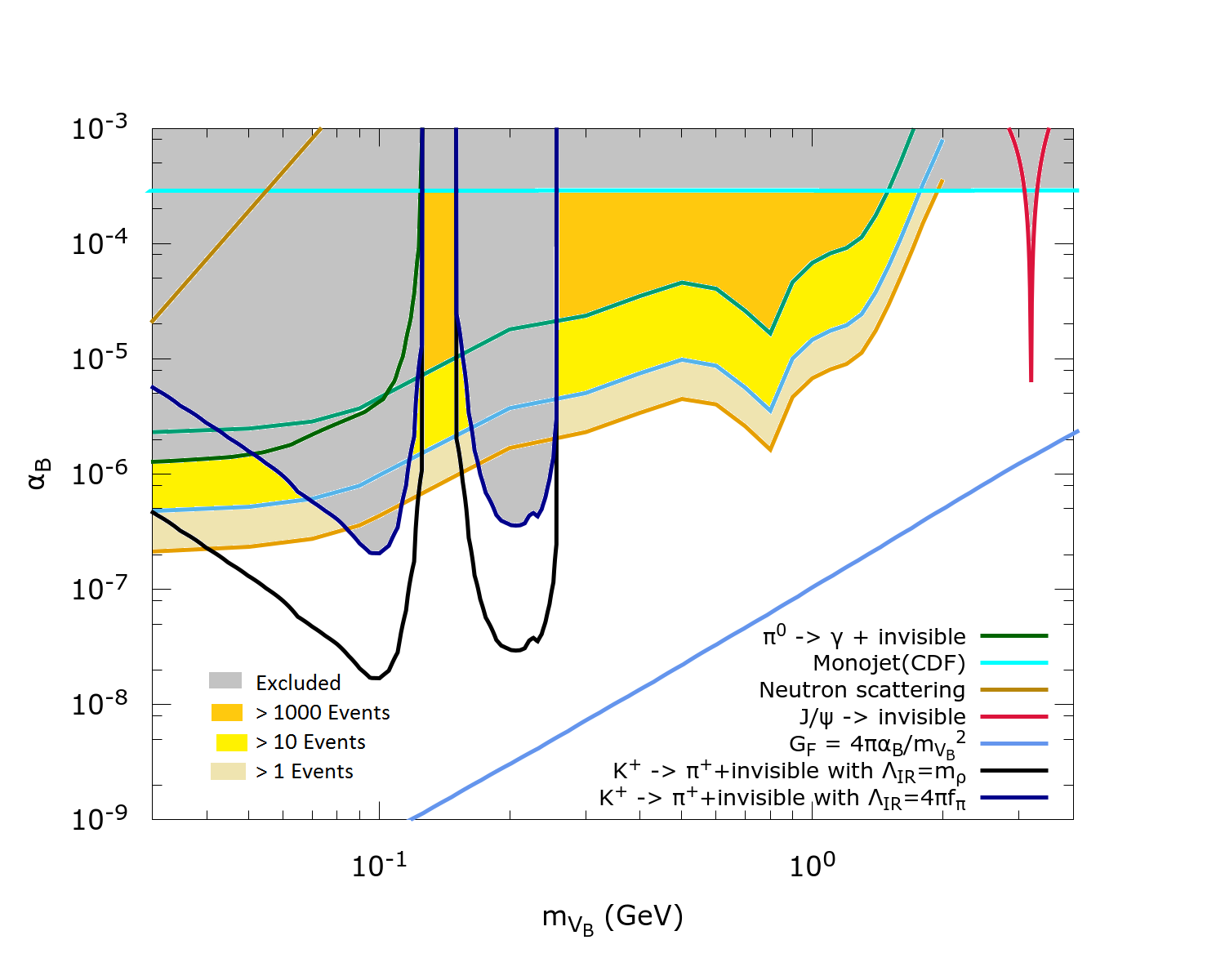}
\caption{The contour sensitivity plot for light dark matter signatures at DUNE experiment in the parameter space of $\alpha_{B} - m_{V_{B}}$.
These dark matter are produced from distinct channels by using 120 GeV proton beam. Here we have considered $m_{\chi} = 0.01$ GeV,
$\epsilon$ = $0$ and POT = $1.1 \times 10^{21}$. In above plot,the gray regions are excluded by existing constraints, while the yellow
contours indicate 1, 10 and 1000 events.}
\label{fig1}
\end{figure}

\begin{figure}[H]
\centering
\includegraphics[width=1.0\linewidth]{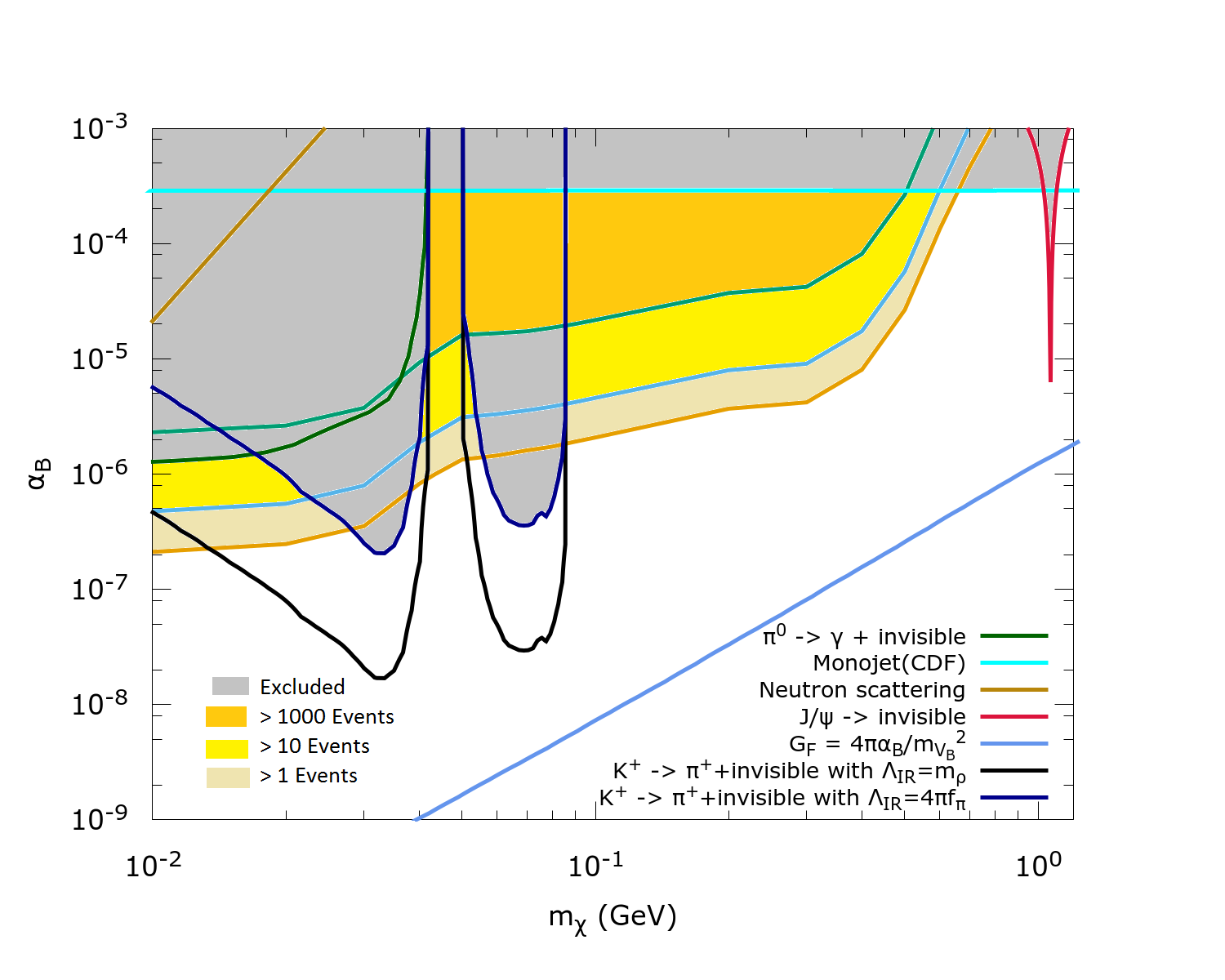}
\caption{The contour sensitivity plot for light dark matter signatures at DUNE experiment in the parameter space of $\alpha_{B} - m_{\chi}$.
These dark matter are produced from distinct channels by using 120 GeV proton beam. Here we have considered $m_{V_{B}} = 3m_{\chi}$ GeV,
$\epsilon$ = $0$ and POT = $1.1 \times 10^{21}$. In above plot,the gray regions are excluded by existing constraints, while the yellow
contours indicate 1, 10 and 1000 events.}
\label{fig3}
\end{figure}
In the figure [\ref{fig2}] we have explored the nature of DM-nucleon scattering event with varying mass of the vector boson mediator
$m_{V_{B}}$ for a fix value
of DM mass i.e. 10 MeV. From this piece of work we observe that meson decay channel is the dominant channel for the production of DM candidates till
600 MeV mass of $V_{B}$.
For intermediate mass range of $V_{B}$ i.e. from 40 MeV to 1 GeV the DM particles are dominantly produced via resonance decay channel
whereas parton-level DM production channel is active for complete range of 
$m_{V_{B}}$ but for $m_{V_{B}} > 1$ GeV the DM production is mainly due to this channel only.

In the figure [\ref{fig1}] we have presented the sensitivity of DUNE for leptophobic DM scenario in $m_{V_{B}}-\alpha_{B}$ parameter space
where $m_{V_{B}}$ is allowed to vary between 0.03 GeV to 2 GeV and baryonic fine structure constant $\alpha_{B}$ varies from $10^{-9}$ to
$10^{-3}$.
This study is performed for the DM mass 10 MeV and kinetic mixing parameter $\epsilon$ = 0. 
The DM-nucleon elastic scattering event contours for 1, 10 and 1000 events are shown
in the figure [\ref{fig1}] along with other experimental constraints.
This plot shows that DUNE experiment can prob DM for baryonic fine 
structure coupling below the existing bound on the
coupling value $\alpha_{B} \simeq 10^{-6}$. 
This lower coupling values are present
for mass of vector boson mediator $V_{B}$ less than 200 MeV.

The figure [\ref{fig3}] shows the contour plot of DM-nucleon elastic scattering events in $m_{\chi}-\alpha_{B}$ plane
where the variation of $m_{\chi}$ is from 0.01 GeV to 1 GeV and $\alpha_{D}$ is varied from $10^{-9}$ to $10^{-3}$. 
This study is performed for 
kinetic mixing parameter $\epsilon$ = 0 and vector boson mass $m_{V_{B}} = 3m_{\chi}$.
The three contours of figure [\ref{fig3}] represents the 1, 10 and 1000 events in the parameter space of $m_{\chi}-\alpha_{B}$.
In this study we observe that the vector boson mediator $V_{B}$ shows weak coupling (below the existing limit of $\alpha_{B} = 10^{-6}$) 
for DM masses less than 250 MeV.

In all three plots we notice a sharp peak around $m_{V_{B}} = 3m_{\chi} \sim m_{\omega} \sim 800$ MeV which can be attributed to the resonance
production via bremsstrahlung process at this point.


\section{Conclusion:}

In our work DUNE simulation study is performed for leptophobic light DM signatures via via DM-nucleon elastic scattering in the parameter space of
$m_{V_{B}}-\alpha_{B}$ and $m_{\chi}-\alpha_{B}$.
The study is based on the benchmark DM model in which DM candidates dominantly interacts with quarks via vector boson mediator $V_{B}$ 
of baryonic gauge group
$U(1)_{B}$. The analysis of figure [\ref{fig1}] and [\ref{fig3}] illuminates the DUNE experiment sensitivity for
leptophobic light DM imposes constraints on the coupling value $\alpha_{B}$, which is lower than the present constraint value of the coupling 
$\alpha_{B} \simeq 10^{-6}$.
These lower coupling value can be explored by DUNE for vector boson mediator mass $m_{V_{B}} < 200$ MeV. Therefore DUNE experiment in beam dump mode
will be able to provide the new results for leptophobic DM.
These results will help us to understand the nature of DM and its interactions.

\begin{acknowledgement}
We are thankful to Prof. Raj Gandhi for his valuable help at all stages of work and for his comments on the manuscript.

\end{acknowledgement}\vspace{-10mm}
%


\end{document}